Calibrating the atomic balance by carbon nanoclusters

Fengqi Song1,*, Xuefeng Wang2, Rebecca C. Powles3, Longbing He[1], Nigel A. Marks4, Shifeng Zhao1, Jianguo Wan[1], Zongwen Liu3, Jianfeng Zhou1, Simon Ringer3, Min Han1, Guanghou Wang[1]

1National Laboratory of Solid State Microstructures, Nanjing University, 210093, Nanjing, China

2Department of Physics and Jiangsu Key Laboratory of Thin Films, Soochow University, 215006, Suzhou, China

3Australian Key Centre for Microscopy and Microanalysis, the University of Sydney, NSW 2006, Australia

4Nanochemistry Research Institute, Curtin University of Technology, Perth, WA 6845, Australia

* Corresponding Author: Fax +86-25-83595535, e-mail: songfengqi@nju.edu.cn

Abstract: Carbon atoms are counted at near atomic-level precision using a scanning transmission electron microscope calibrated by carbon nanocluster mass standards. A linear calibration curve governs the working zone from a few carbon atoms up to 34,000 atoms. This linearity enables adequate averaging of the scattering cross sections, imparting the experiment with near atomic-level precision despite the use of a coarse mass reference. An example of this approach is provided for thin layers of stacked graphene sheets. Suspended sheets with a thickness below 100 nm are visualized, providing quantitative measurement in a regime inaccessible to optical and scanning probe methods.

Counting at atomic-level precision is one of ultimate goals of nanometrology[1,2,3-6,7]. Quantitative scanning transmission electron microscopy (Q-STEM) is characterized [2,3-5,8] by simultaneous determination of nanometer-resolution lateral morphology and high sensitivity to the number of local atoms, N. Using Q-STEM, metrology of N can be directly performed on the STEM images without any image simulation and even generates their mass spectra [5,7,9]. Experimental precision down to several atoms has recently been demonstrated with Q-STEM using gold clusters [5]. This approach attains similar precision to that provided by custom-designed nanomechanical mass sensors [6]. However, Q-STEM atomic balance is presently limited to heavy atoms, even if the imaging resolution of individual low-Z atoms has been reported [10].

Calibration is the crucial step in implementing the atomic-level precision. This process primarily consists of seeking mass standards with determined N and carefully improving the signal noise ratio (SNR) from the tiny nanoobjects. In atom counting by Q-STEM, the quantity of incoherently scattered electrons is monotonically dependent on N [3,11]. A high angle annular dark field (HAADF) detector records the scattered electrons and forms the image with quantitative information [3]. The main work then lies in constructing the calibration curve of the HAADF intensities against N and determining its slope for small N, i.e. the constant of cross sections of electron scattering and collection ($C_{con}$). In the absence of accurate theoretical calculation of $Cc_{on}$[11], a practical rule is that the HAADF intensities have to be calibrated for each species of atom [5]. Here poor SNR is a common problem and the ultimate atomic precision is always extracted very much close to the noise limit of the experimental configuration. With gold, for example, errors of 20% are typical for nanoclusters with hundreds of atoms and even higher errors are present for smaller clusters3,5. Poorer SNRs are expected for low-Z nanoclusters with the same N due to reduced electron scattering4. The HAADF intensities of small low-Z clusters with tens of atoms on an amorphous support normally approach the level of background noise 4, and so on one hand, the poor SNR requires clusters with larger N in calibration of low-Z atoms. On the other hand, the calibration curve is nonlinear for larger $N$. A logarithmic relation is present for gold clusters with more than 1500 atoms 5. The conflict between the two main tasks imposes a dilemma in building an atomic balance of low-Z structures, which relies on the experimental determination of the calibration curve over an extended range of values of N. Here we show its implementation by the calibration of carbon nanoclusters. A linear calibration curve enables near atomic-level precision of the approach. Suspended few-layer graphene (FLG) sheets are counted as a demonstration.

To build ultrafine calibration references for use as atomic-level mass standards we first require a series of carbon nanoparticles with known N. Such requirements are satisfied by particles collected from a carbon cluster beam generated in a radiofrequency sputtering chamber of a gas-aggregation cluster source 12. Argon flow rates of 91 sccm and Helium flow rates of 22 sccm were used, and sputtering was performed at pressures up to 200 Pa with a net input power of 300 W. Note that liquid nitrogen cooling, normally used along the drift zone, was not applied, resulting in a room-temperature cluster beam. The amorphous nature of the carbon clusters are demonstrated by high-resolution TEM, electron diffraction and electron energy loss spectra (EELS), which eliminate the influence of Bragg scattering in electron collection. For clusters with diameters of 3.5-8 nm, quantum mechanical effects dependent on N are of little relevance 13, and spherical shapes are energetically favored, particularly for the room-temperature beams emerging from the gas aggregation chamber 13. In other words, the carbon clusters are spherical when they are moving within the beam, and the value of N for the clusters can be calculated from their spherical shapes.

To confirm the stability of the spherical shape when depositing the clusters, we performed molecular dynamics (MD) simulations. Amorphous carbon clusters comprise multiple hybridization states which may potentially change during impact. Accordingly, we employ the environment-dependent interaction potential (EDIP) for carbon 14, which has been extensively applied to amorphous carbon in simulations of quenching and deposition15. In Fig. 1a, an amorphous $C_{1441}$ cluster is shown being deposited onto a diamond substrate containing 23,120 atoms at a kinetic energy of 0.25 eV/atom. The spherical shape is maintained as shown in Fig. 1b and the supplementary movie. It confirms the stability of the spherical shapes and the density of the amorphous carbon clusters upon nearly-free deposition. Therefore, free deposition conditions, i.e. no accelerating potential, were used to deposit the reference clusters for calibration. It is shown that the shape evolution of clusters on collision with the substrate is dependent on the impact velocity when the cluster and substrate are composed of the same material 13,16. Clusters with different sizes from the same supersonic beam have similar velocities 17. Free deposition is believed to guarantee the soft landing of the clusters with little transformation as confirmed by the simulation. The above evidence supports our assumption that all of the references keep their spherical shape after impact onto the substrate using free deposition conditions. Under this assumption, the lateral diameter of the deposited nanoclusters is equal to their vertical height. Their atom counts N can be calculated based on the cluster geometry and density as determined by EELS.

Secondly, the reference clusters are sampled to extract the HAADF contributions, where the intensity from the support film sets the zero level (Fig 2a). TEM grids without carbon coating (XXBR) were used[18]. The portion above the zero level is sampled as the HAADF contribution from the nanocluster. We select a small region in the center of the particle and average the intensity of each pixel, giving the HAADF maximum. We sum the intensities of all the pixels from a nanocluster and quote it as the HAADF integral.

Finally, the HAADF maximum and HAADF integral are plotted against the diameter of the reference clusters in Fig 2b, while the HAADF integral is plotted against N in Fig. 2c. All curves increase monotonically, even though the atomic columns are unresolved by the TEM. The linearity up to 8 nm [equivalently $(3.4 \pm 0.6) \times 10^4$ atoms] is shown in both the HAADF maximum curve (Fig. 2b) and the calibration curve (Fig. 2c). Note that the cluster diameters are much larger than the size of the electron probe (< 0.8 nm), and so the integral HAADF intensities sum the HAADF values of all the probed points over a selected cluster, with the number of probed points increasing with the diameter of the cluster. As the result, the integral HAADF curve in Fig. 2b depends on the projected area of the cluster and increases in a polynomial fashion in contrast to the linear increase of the HAADF maximum intensity curve. In Fig. 2b, the levels of statistical noise and the HAADF maximum of the smallest reference cluster are indicated by arrows respectively; the former determines the sensitivity of the technique.

The calibration curve demonstrates that this metrology technique can achieve near atomic-level precision. The slope of the curve in Fig. 2c is the cross section constant $C_c$ on, and the linear portion extends to atom counts of very few atoms. In addition, the linear fitting is essentially the average of $C_c$ on over the whole linear region, further reducing measurement errors arising from the coarse determination of N and HAADF intensities. This results in a determined $C_{con}$ with an error of 19%, which is valid even for small N.

In a demonstration of atomic-level metrology, we consider layer counting of suspended FLG sheets. Such studies are of particular importance due to the fact that the substrate critically influence the electronic structures of a contacted graphene monolayer 19,20.. Note that layer counting by scanning probe microscopy is not suitable for suspended FLG membranes due to the delicate force balance between the tip and suspended FLG films, while optical approaches such as Raman and color contrast 21 fail lack lateral resolution at the nanoscale. Applying the calibration curves above, the numbers of the layers (L) were counted in a graphene sheet suspended on a porous carbon film (Fig. **3**a). The cleaved graphene sheet is composed of several regions of 3L, 10L, 28L and 35L. The lateral extent of the 3L block is less than 100 nm, which is invisible to normal optical techniques. In order to evaluate the layer resolution, all the HAADF signals over the regions of vacuum, 3L, 10L, and 28L were extracted for statistics, yielding a histogram curve (see blue curve in Fig. 3b). The full width at half maximum (FWHM) decreases with decreasing layer number, and results of 3 ± 1.3L and 10±1.5L are obtained. The error ratio is further reduced in measuring sheets with larger L, e.g. 28±2.6L. The number of atoms probed by the STEM sampling can calculated by considering a cylindrical nano-block with the diameter of the electron probe and the height of the specimens22. In the 3L graphene measurement with the determined $C_{con}$, the probe with the diameter of 0.8nm counts the minimum nano-block of 54±23 carbon atoms. We also measured a 1L sheet with a more focused probe (0.5 nm diameter), corresponding to a precision of 5.3 carbon atoms. This further confirms the near atomic-level precision of the STEM balance. Even better precision is expected in an ideal combination of highly-focused beam, low-noise detector and ultrathin and uniform support 20.

To conclude, HAADF intensities from carbon atoms have been calibrated and their linearity has been demonstrated to be valid up to 34,000 atoms. We demonstrate the technique by measuring thicknesses of nanoscale regions of graphene with thicknesses down to one monolayer. Near atomic-level precision is demonstrated in measurements on FLG sheets. The approach is applicable to other low-Z materials.

This work was financially supported by the National Natural Science Foundation of China (Grant Nos. 90606002, 10674056, 10775070 and 10904100), the National Key Projects for Basic Research of China (Grant No. 2009CB930501, 2010CB923401), and the Program for New Century Excellent Talents in University of China (Grant No. NCET-07-0422). We thank Prof. Xiaoning Zhao for their technical assistance.

with the sub-nanometer diameter and a 4nm-height carbon atom column. The electron probe is no bigger than 0.8nm. Therefore`, N in the column is (3.14159*4nm*0.4nm*0.4nm*138 atom/nm3 ~277atoms) at most. Considering the error of HAADF determination of 31/308 and the error of 28% from atom number calculation`, our finest counterpoise is made of 277 atoms with the error of 212atoms. The probe size is estimated by the STEM imaging of a film with small Pb clusters`, when only the clusters larger than 0.75nm are visible.

Figure Captions

Figure 1. MD simulation of free deposition of amorphous carbon clusters. (a) and (b). Snapshots of a cluster of C1441 before incidence and after collision, respectively. The simulation time is 0.388 and 2.294ps. The spherical shape was well maintained after the deposition of 0.25 eV/atom. The colors (blue, green and red) of the atoms represent atoms with sp, sp2 and sp3 hybridization respectively. (enhanced online)

Figure 2. The calibration curves. (a) Extraction of a reference cluster of 6nm. (b) The HAADF intensities against the particles' diameters. The upper curve (black) in (b) shows the HAADF maximum of the sampling clusters and the other curve shows the integral HAADF intensities over all the sampling points of a selected cluster. The EELS measurement gives an sp3 ratio of 52% with the error of 20%, therefore the density of 2.77g/cm3 is used for the nanoclusters (see reference 25). N is then calculated as shown in (c), which gives a linear calibration curve of the integral HAADF intensities plotted against N.

Figure 3. Measuring a suspended few-layer graphene (FLG) sheet. (a) The FLG sheet suspended by a porous carbon. The darker part is a hole (vacuum region). The layer numbers are marked with 3L, 10L, 28L and 35L. As marked by the boxes, the HAADF signals over the region of vacuum, 3L, 10L, and 28L are extracted for statistics over the HAADF intensities, yielding a histogram of the blue curve in (b). The red line is the result of Gaussian peak fitting. Based on the ideal Gaussian fitting of the red curve, we subtracted the FWHM of the vacuum peak from that of the FLG sheets in order to show the possible improvement by detector optimization. This gives the green peaks inserted below.

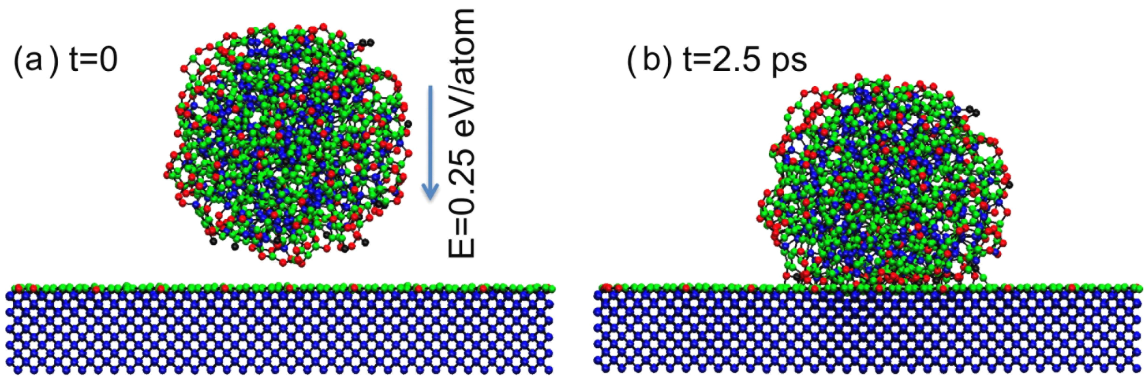

Fig 1

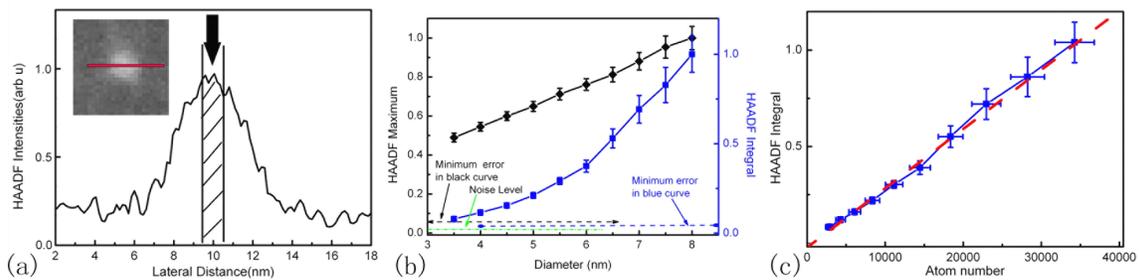

Fig 2

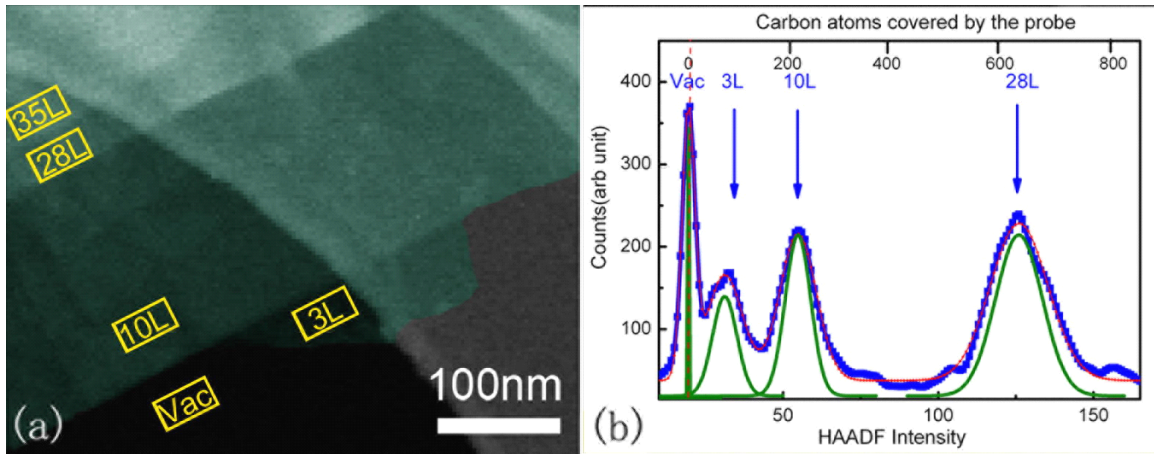

Fig 3